\documentclass[apj, twocolumn, revtex4, 8pt]{emulateapj-rtx4}

\usepackage{color}
\usepackage[bookmarks=false]{hyperref}

\slugcomment{accepted for publication in ApJ}

\shorttitle{Demographics of Ly$\alpha$ Emitter Structures}
\shortauthors{T.Shibuya et al. }

\begin{document}

\title{What is the Physical Origin of Strong Ly$\alpha$ Emission?\\
I. Demographics of Ly$\alpha$ Emitter Structures \altaffilmark{\ddag}}


\author{Takatoshi SHIBUYA\altaffilmark{1,2}, Masami OUCHI\altaffilmark{1,3}, Kimihiko NAKAJIMA\altaffilmark{3,4}, Suraphong YUMA\altaffilmark{1}, Takuya HASHIMOTO\altaffilmark{3}, Kazuhiro Shimasaku\altaffilmark{3,5}, Masao MORI\altaffilmark{2}, and Masayuki UMEMURA\altaffilmark{2}}
\email{shibyatk\_at\_icrr.u-tokyo.ac.jp}

\altaffiltext{1}{Institute for Cosmic Ray Research, The University of Tokyo, 5-1-5 Kashiwanoha, Kashiwa, Chiba 277-8582, Japan}
\altaffiltext{2}{Center for Computational Sciences, The University of Tsukuba, 1-1-1 Tennodai, Tsukuba, Ibaraki 305-8577 Japan}
\altaffiltext{3}{Department of Astronomy, Graduate School of Science, The University of Tokyo, Tokyo 113-0033, Japan}
\altaffiltext{4}{Institute for the Physics and Mathematics of the Universe (IPMU), TODIAS, The University of Tokyo, 5-1-5 Kashiwanoha, Kashiwa, Chiba 277-8583, Japan}
\altaffiltext{5}{Research Center for the Early Universe, Graduate School of Science, The University of Tokyo, Tokyo 113-0033, Japan}

\altaffiltext{\ddag}{Based on data obtained with the Subaru Telescope operated by the National Astronomical Observatory of Japan.}


\begin{abstract}

We present the results of structure analyses for a large sample of $426$ Ly$\alpha$ emitters (LAEs) at $z\sim2.2$ that are observed with {\it HST}/ACS and WFC3-IR by deep extra-galactic legacy surveys. We confirm that the merger fraction and the average ellipticity of LAE's stellar component are $10-30$\% and $0.4-0.6$, respectively, that are comparable with previous study results. We successfully identify that some LAEs have a spatial offset between Ly$\alpha$ and stellar-continuum emission peaks, $\delta_{\rm Ly\alpha}$, by $\sim 2.5-4$ kpc beyond our statistical errors. To uncover the physical origin of strong Ly$\alpha$ emission found in LAEs, we investigate Ly$\alpha$ equivalent width (EW) dependences of these three structural parameters, merger fraction, $\delta_{\rm Ly\alpha}$, and ellipticity of stellar distribution in the range of $EW({\rm Ly\alpha})=20-250$\,\AA. Contrary to expectations, we find that merger fraction does {\it not} significantly increase with Ly$\alpha$ EW. We reveal an anti-correlation between $\delta_{\rm Ly\alpha}$ and $EW({\rm Ly\alpha})$ by Kolmogorov-Smirnov (KS) test. There is a trend that the LAEs with a large Ly$\alpha$ EW have a small ellipticity. This is consistent with the recent theoretical claims that Ly$\alpha$ photons can more easily escape from face-on disks having a small ellipticity, due to less inter-stellar gas along the line of sight, although our KS test indicates that this trend is not statistically significant. Our results of Ly$\alpha$-EW dependence generally support the idea that an H {\sc i} column density is a key quantity determining Ly$\alpha$ emissivity.

\end{abstract}

\keywords{cosmology: observations --- early universe --- galaxies: formation --- galaxies: high-redshift}


\section{INTRODUCTION}

Ly$\alpha$ Emitters (LAEs) are a population of high-$z$ star-forming galaxies selected with a narrow-band (NB) and broad-band (BB) filters to identify their prominent Ly$\alpha$ emission. A large number of NB observations have been carried out to study LAEs at $z\sim3-7$ and beyond $z=7$ \citep[e.g., ][]{2010ApJ...711..928C,2007ApJ...667...79G,2012ApJ...744..110C,2008ApJS..176..301O,2010ApJ...723..869O,2010ApJ...725..394H,2007ApJ...660.1023F,2011ApJ...734..119K,2006ApJ...648....7K,2012ApJ...752..114S}. Based on high resolution imaging and spectral energy distribution (SED) fitting, such a galaxy population is thought to be typically young, compact, less-massive, less-dusty, and a possible progenitor of Milky Way mass galaxies \citep[e.g., ][]{2011ApJ...743....9G,2011ApJ...733..114G,2010MNRAS.402.1580O,2007ApJ...671..278G,2011ApJ...740...71D,2008ApJ...681..856R}. LAEs are also used as a powerful probe for estimating the neutral hydrogen fraction at the reionizing epoch, because Ly$\alpha$ photons are absorbed by intergalactic medium (IGM). 

Despite the significant importance of LAEs in galaxy formation and cosmology, Ly$\alpha$ emitting mechanism is not completely understood due to the highly-complex resonant nature of Ly$\alpha$ in the interstellar medium (ISM). Many theoretical models have predicted that the neutral gas and/or dust distributions surrounding central ionizing sources are closely related to the escape of Ly$\alpha$ \citep[e.g., ][]{2013ApJ...766..124L,2009ApJ...704.1640L,2007ApJ...657L..69L,2013arXiv1302.7042D,2013arXiv1308.1405Z,2010ApJ...716..574Z,2012arXiv1209.5842Y}. Resonant scattering in neutral ISM results in a significant attenuation of Ly$\alpha$. These Ly$\alpha$ absorbing ISM may be blown out by galaxy mergers and subsequent galactic outflows. The galactic interactions would also trigger the star formation and enhance Ly$\alpha$ emissivity. In fact, merging features have been found in the LAE population in several observational studies \citep[e.g., ][]{2009ApJ...705..639B,2010ApJ...716L.200B,2012ApJ...753...95B, 2011ApJ...743....9G}. These studies have investigated various morphological properties for $\sim100-200$ LAEs at $z\sim2-3$, but have not examined dependences on Ly$\alpha$ emissivity. 

It is also informative to investigate Ly$\alpha$ emitting positions relative to star forming regions. The geometry of surrounding neutral gas might leave an imprint on the spatial offsets between Ly$\alpha$ and stellar-continuum emission. \citet{2013ApJ...773..153J} have investigated the spatial offsets for $\sim70$ LAEs at $z=6-7$, and found misalignments in several objects. However, they have not studied systematically the spatial offsets and its dependence on physical properties of LAEs. 

In addition, the galactic morphologies are considered to be relevant to the Ly$\alpha$ emissivity. Copious amounts of gas and/or dust are likely to inhabit in the galactic disk. Consequently, Ly$\alpha$ photons preferentially escape out in the direction perpendicular to the disk. The inclination effect on Ly$\alpha$ emissivity has been widely examined theoretically \citep[e.g., ][]{2013arXiv1308.1405Z,2012A&A...546A.111V,1993ApJ...415..580C,1994ApJ...432..567C,2011MNRAS.416.1723B,2007ApJ...657L..69L,2009ApJ...704.1640L,2010ApJ...716..574Z}. These studies have predicted the preferential escape of Ly$\alpha$ in the face-on direction. 

However, these structural properties and their dependences on Ly$\alpha$ emissivity have not yet been examined statistically for high-$z$ LAEs. Ly$\alpha$ emissivity is tightly related with Ly$\alpha$ EW, since the EW represents a Ly$\alpha$ luminosity normalized by star formation activity in a galaxy. A systematic study of the relation between the structures of LAEs and their Ly$\alpha$ EW will provide crucial hints of the neutral gas distributions and associated Ly$\alpha$ emitting mechanisms. 


This is the first paper in the series exploring the Ly$\alpha$ emitting mechanisms\footnote{The second paper presents a kinematic study for LAEs \citep{2014arXiv1402.1168S}.}. In this paper, we present results of our study on structures of $z\sim2.2$ LAEs to verify the Ly$\alpha$-EW dependence of merger fraction, Ly$\alpha$ spatial offset $\delta_{\rm Ly\alpha}$, and ellipticity by exploiting the {\it Hubble Space Telescope} ({\it HST})/ACS and WFC3 images. We use our statistically-large sample consisting of $\sim3400$ LAEs constructed with Subaru NB observations. First we describe details of our $z=2.2$ LAE sample for our structure analyses in \S \ref{sec_samples}. Next, we explain our methods to derive structural quantities in the rest-frame UV/optical emission in \S \ref{sec_methodology}. We examine the dependence of the derived morphological quantities on Ly$\alpha$ EW in \S \ref{sec_results}. In \S \ref{sec_discussion}, we discuss physical mechanisms by which high-$z$ galaxies emit Ly$\alpha$. In the last section \S \ref{sec_conclusion}, we summarize our findings. 

Throughout this paper, we adopt the concordance cosmology with $(\Omega_m, \Omega_\Lambda, h)=(0.3, 0.7, 0.7)$, \citep{2011ApJS..192...18K}. All magnitudes are given in the AB system \citep{1983ApJ...266..713O}.

\section{SAMPLE}\label{sec_samples}

Our LAE sample for the structure analysis has been constructed by observations with Subaru/Suprime-Cam \citep{2002PASJ...54..833M} equipped with the NB filter, NB387 ($\lambda_c = 3870$ \AA\, and FWHM $= 94$ \AA) \citep{2012ApJ...745...12N}. The Suprime-Cam observations have been carried out for LAEs at $z=2.2$ with NB387 in a total area of $\sim1.5$ square degrees. Based on the color selection of $B-NB387$ and $u^*-NB387$, the Suprime-Cam observations have located $619$, $919$, $747$, $950$, and $168$ LAEs in the Cosmic Evolution Survey (COSMOS) \citep{2007ApJS..172....1S}, the Subaru/{\it XMM-Newton} Deep Survey (SXDS) \citep{2008ApJS..176....1F}, the Chandra Deep Field South (CDFS) \citep{2001ApJ...551..624G}, the {\it Hubble} Deep Field North (HDFN) \citep{2004ApJ...600L..93G}, and the SSA22 \citep[e.g., ][]{2000ApJ...532..170S} fields, respectively. In the above five fields, a total of $\sim3400$ LAEs have been selected down to a Ly$\alpha$ EW of $20-30$ \AA\, in rest-frame (Nakajima et al. in prep.). This large sample size enables us to study statistically various properties of high-$z$ LAEs, such as their metal abundances \citep{2012ApJ...745...12N, 2013ApJ...769....3N}, Ly$\alpha$ velocity offset \citep{2013ApJ...765...70H,2014arXiv1402.1168S}, and Ly$\alpha$ halo (Momose et al. in prep.). Details of observations and selection for LAEs are presented in \citet{2012ApJ...745...12N, 2013ApJ...769....3N}.

\section{METHODOLOGY}\label{sec_methodology}

In this section, we describe methods of our structure analysis using the {\it HST} data. We focus mainly on three structural properties: the merger fraction (\S \ref{subsec_method_merger}), the spatial offset between Ly$\alpha$ and stellar-continuum positions $\delta_{\rm Ly\alpha}$ (\S \ref{subsec_method_offset}), and the ellipticity (\S \ref{subsec_method_ellipticity}). 

We use the $I_{814}$ and $H_{160}$ data taken with Advanced Camera for Surveys (ACS) and Wide Fields Camera 3 (WFC3) on {\it HST}, respectively, to examine the rest-frame UV and optical morphology of the LAE counterparts. The COSMOS, SXDS, and GOODS-North and South fields are partially imaged by the Cosmic Assembly Near-infrared Deep Extragalactic Legacy Survey \citep[CANDELS; ][]{2011ApJS..197...35G, 2011ApJS..197...36K} with {\it HST}/ACS and WFC3. The $5\sigma$ limiting magnitudes in a $0.\!\!^{\prime\prime}2$ aperture are $28.3-29.4$ in $I_{814}$ and $26.5-27.6$ in $H_{160}$. Additionally, the COSMOS field is mostly covered by the ACS imaging with the $I_{814}$ filter \citep[COSMOS-Wide; ][]{2007ApJS..172..196K, 2010MNRAS.401..371M}; however its depth is $\sim1-2$ magnitudes shallower than that of CANDELS. We use both of the CANDELS and COSMOS-Wide fields for our morphological analysis. The typical sizes of the point spread function (PSF) are $0.\!\!^{\prime\prime}09$ and $0.\!\!^{\prime\prime}18$ ($\sim0.75$ and $\sim1.5$ kpc at $z=2.2$) in the $I_{814}$ and $H_{160}$ images, respectively. The number of LAEs in the {\it HST} fields are summarized in Table \ref{table_hst}. 

\begin{deluxetable*}{ccc}
\tabletypesize{\scriptsize}
\tablecaption{Number of Our Ly$\alpha$ Emitters}
\tablehead{  \colhead{Field} & \colhead{$m_{\rm lim}$} & \colhead{Number of LAEs} \\ 
\colhead{(1)}& \colhead{(2)}& \colhead{(3)} } 

\startdata
    & $I_{814}$ on ACS & \\ \hline
    COSMOS-Wide & $27.0$ & 564 \\ 
    CANDELS GOODS-S & $28.5$ & 213 \\ 
    CANDELS GOODS-N & $28.3$ & 95 \\ 
    CANDELS UDS & $28.4$ & 70 \\ \hline
    Total Number & & 942 \\
    & $H_{160}$ on WFC3 & \\ \hline
    CANDELS COSMOS & $26.9$ & 86 \\ 
    CANDELS GOODS-S & $26.7$-$27.6$ & 65 \\ 
    CANDELS GOODS-N & $26.5$ & 63 \\ 
    CANDELS UDS & $27.1$ & 83 \\ \hline
    Total Number & & 297 
\enddata
\tablecomments{Columns: (1) Field. (2) $5\sigma$ limiting magnitude in a $0.\!\!^{\prime\prime}2$ aperture. (3) Number of $z=2.2$ LAEs taken with {\it HST}.}
\label{table_hst}
\end{deluxetable*}

\subsection{Identifications of LAE Counterparts}\label{subsec_cutout}

In order to search for UV and optical counterparts of our LAEs, we first extract $3^{\prime\prime}\times3^{\prime\prime}$ cutout images from the $I_{814}$ and $H_{160}$ data at the position of each LAE in the similar manner as previous morphological studies \citep{2009ApJ...705..639B,2010ApJ...716L.200B,2012ApJ...753...95B, 2011ApJ...743....9G}. The size of cutouts is exactly the same as \citet{2012ApJ...753...95B} who have studied morphology of $z=2.1$ LAEs. In total, we obtain $942$ and $297$ cutout images of $I_{814}$ and $H_{160}$ bands, respectively. 

Next, we detect sources in the {\it HST} cutout images, and perform photometry for all of the sources having an area larger than five contiguous pixels ({\tt DETECT\_MINAREA} $= 5$) with a flux greater than $2.5\sigma$ over the sky surface brightness ({\tt DETECT\_THRESH} $= 2.5$) using {\tt SExtractor} version 2.8.6 \citep{1996A&AS..117..393B}. Our {\tt DETECT\_THRESH} value is higher than that in \citet[; {\tt DETECT\_THRESH} $= 1.65$]{2012ApJ...753...95B}. Because one of our aims is to estimate merger fraction from the number of close galaxy pairs (\S \ref{subsubsec_method_closepair}), a reliable detection of objects is required even for very faint sources. If a low {\tt DETECT\_THRESH} value is chosen, false detections increase, leading to an overestimate of a merger fraction. Here, we tune a {\tt DETECT\_THRESH} parameter to identify real galaxy pairs by visual inspection. Meanwhile, {\tt DETECT\_THRESH} is set to $1.65$ instead of $2.5$ when we derive morphological indices for counterparts of LAEs in \S \ref{subsubsec_method_indices}. 

\begin{figure}[t!]
  \begin{center}
    \includegraphics[width=70mm]{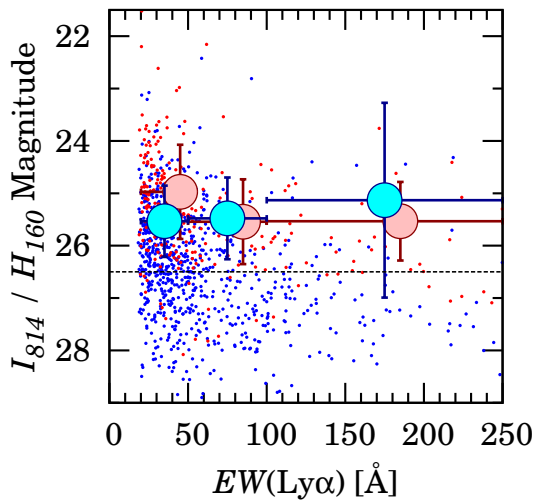}
  \end{center}
  \caption[EW and UV magnitude]{{\footnotesize Ly$\alpha$ EW and $I_{814}$/$H_{160}$ magnitudes of continuum counterparts. Blue and red circles denote counterparts in $I_{814}$ and $H_{160}$ data, respectively. Large cyan and magenta circles are the average $I_{814}$ and $H_{160}$ magnitudes, respectively, in EW bins of $20-50$, $50-100$, and $>100$\,\AA. The large magenta circles are slightly shifted along x-axis for the sake of clarity. The dashed horizontal line represents the magnitude cut for the close-pair method, $m=26.5$. With the magnitude cut, the average continuum magnitudes of EW bins are comparable between the EW bins. }}
  \label{fig_ew_mag}
\end{figure}

Finally, we define $I_{814}$ and $H_{160}$ counterparts of LAEs as objects within a radius of $0.\!\!^{\prime\prime}65$ ($\sim5.4$ kpc at $z=2.2$) from an NB source center, following the definition of \citet{2012ApJ...753...95B}. According to the assessment by \citet{2012ApJ...753...95B}, this selection radius can exclude effectively field sources. Figure \ref{fig_ew_mag} shows the Ly$\alpha$ EW and $I_{814}$/$H_{160}$ magnitudes of the counterparts. Figure \ref{fig_ew_mag} reproduces the Ando effect \citep{2006ApJ...645L...9A} that Ly$\alpha$ EW anti-correlates with continuum magnitudes. We adopt a continuum magnitude cut of $26.5$ mag, and derive the average $I_{814}$/$H_{160}$ magnitudes in each Ly$\alpha$ EW bin of $20-50$, $50-100$, and $>100$\,\AA. These average values are almost constant within a $1\sigma$ error bar between the EW bins. In our analyses, we only use objects with a continuum brighter than $26.5$ mag.

\subsection{Merger Fraction}\label{subsec_method_merger}

We estimate the merger fraction of our LAE sample in two methods, the {\it close-pair method} (\S \ref{subsubsec_method_closepair}) and the {\it morphological index method} (\S \ref{subsubsec_method_indices}). The former is to count the number of resolved sources falling within a specific selection radius. The latter is to classify mergers with the non-parametric morphological indices, $CAS$ \citep{2000ApJ...529..886C,2003ApJS..147....1C}, for the sources that are unresolved in the close-pair method. In a merger process, galaxies first approach each other, and finally undergo coalescence(s). The close-pair method selects merger objects in the approaching phase, and the index method identifies the final coalescence phase. In calculations of morphological indices, all sources in a selection radius are usually considered as a galaxy system, even if they are clearly isolated. Using the morphological indices, we aim to examine whether unresolved single sources are interacting (correspondingly morphologically-disturbed) galaxies or intrinsically-isolated components. The classification with the morphological indices is complementary with the close-pair method which identifies mergers with discrete components. 

\begin{figure}[t!]
  \begin{center}
    \includegraphics[width=60mm]{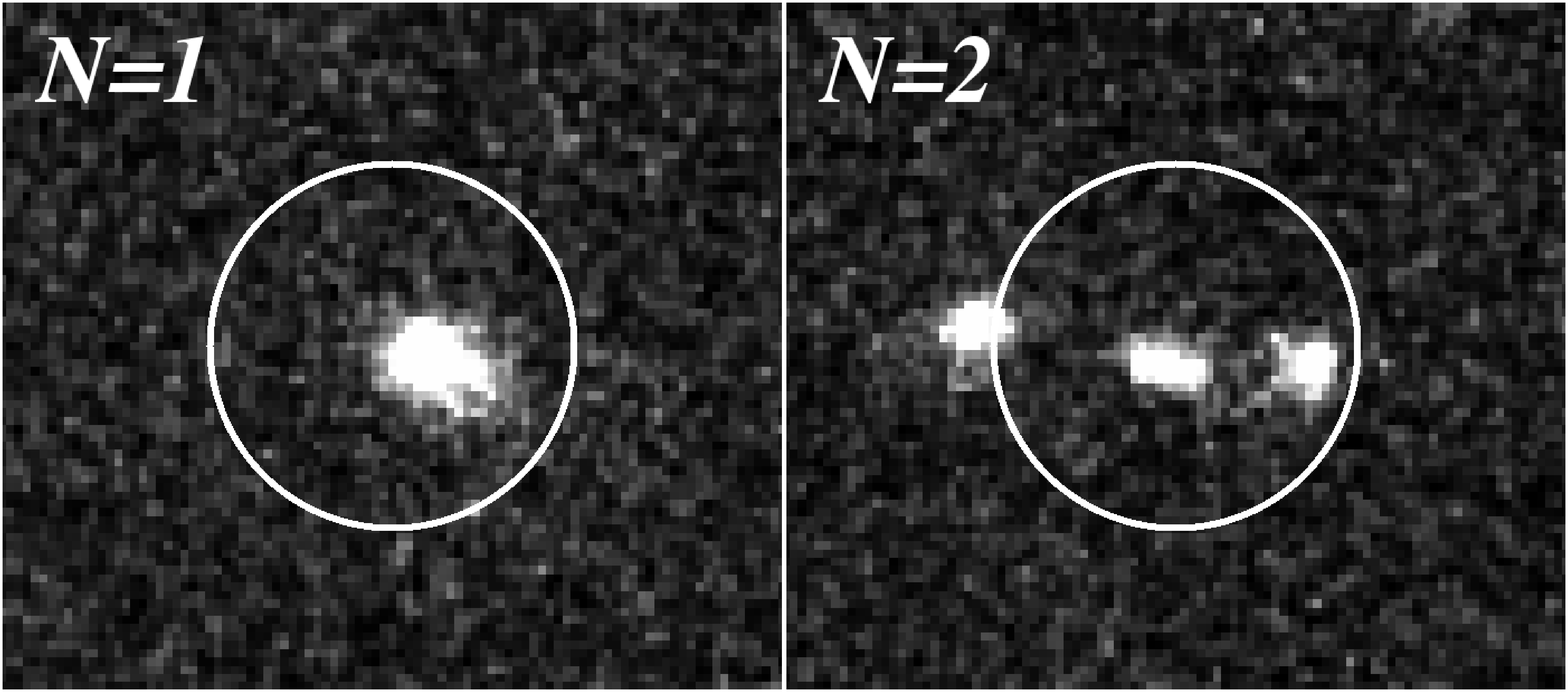}\\
    \includegraphics[width=60mm]{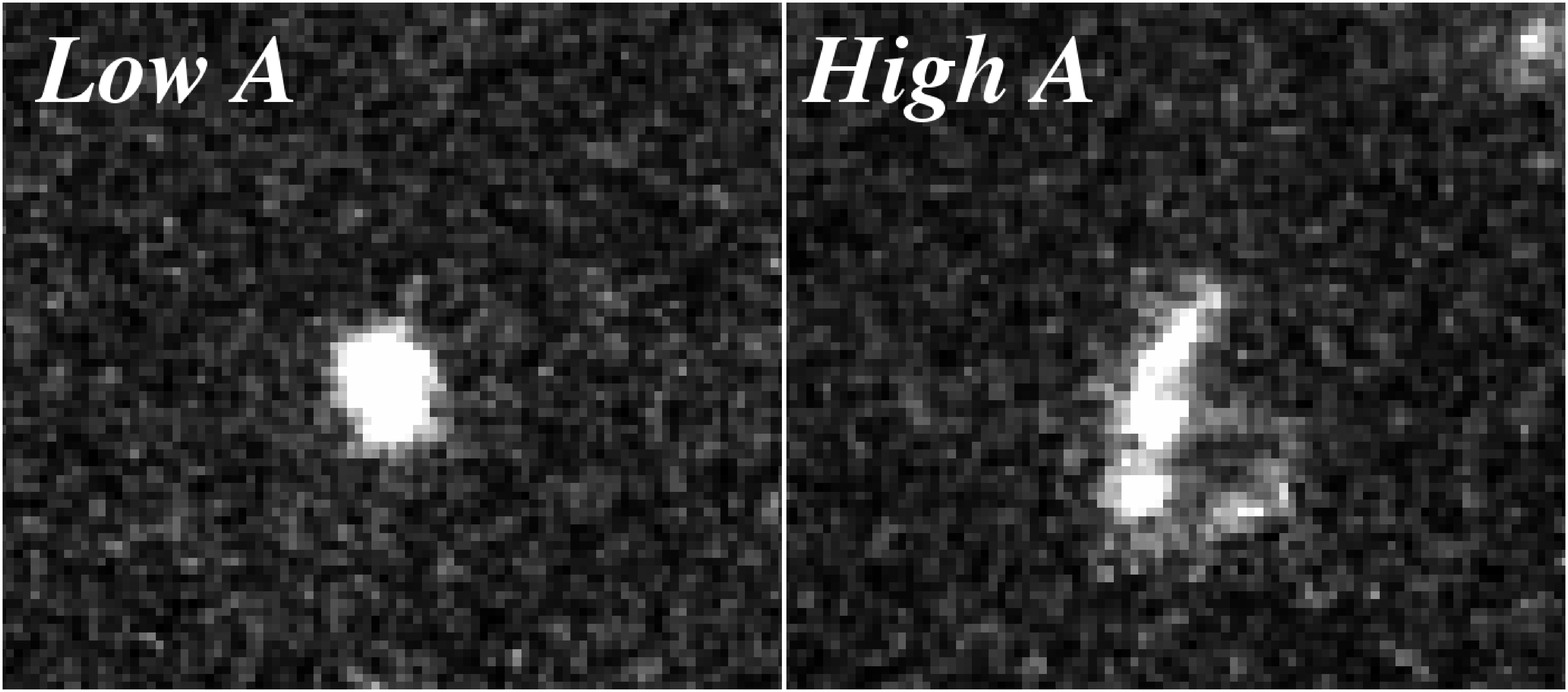}
  \end{center}
  \caption[NB contour]{{\footnotesize Example $I_{814}$ images of mergers and non-mergers in the close-pair (top panels) and the morphological index methods (bottom panels). The UV counterparts in the top-left and right panels show single and multiple components, respectively, within our selection radius of $0.\!\!^{\prime\prime}65$ (white circles). The latter object is classified as a merger in the close-pair method. The UV counterparts in the bottom-left and right panels have a low and high value of asymmetry, respectively. The object with a high $A$ represents a highly-disturbed structure. North is up and east is to the left. }}
  \label{fig_merger_example}
\end{figure}

\subsubsection{Close-Pair Method}\label{subsubsec_method_closepair}

The close-pair method has been used to identify mergers at low- \citep[e.g., ][]{2013MNRAS.435.3627E, 2000MNRAS.311..565L} and high-$z$ \citep[e.g., ][]{2012ApJ...745...85L}. It is extremely difficult to detect faint components of minor mergers at high-$z$. In our analysis, we consider only major mergers with multiple components of comparable flux (a flux ratio of $0.3$--$1$). By counting the number of sources within the selection radius, $r_{\rm sel} = 0.\!\!^{\prime\prime}65$ (\S \ref{subsec_cutout}), we simply define (major) mergers as counterparts with multiple sources. Figure \ref{fig_merger_example} shows representative examples of a merger and a non-merger classified in the close-pair method. 

\begin{deluxetable*}{ccc}
\tabletypesize{\scriptsize}
\tablecaption{Number of Our Ly$\alpha$ Emitters for Each Analysis}
\tablehead{  \colhead{Quantity} & \colhead{Criteria} & \colhead{Number of LAEs} \\ 
\colhead{(1)}& \colhead{(2)}& \colhead{(3)}} 

\startdata
    Merger Fraction & ($I_{814}$$<26.5$) & 426 \\ 
    & ($H_{160}$$<26.5$) & 237 \\ \hline
    Ly$\alpha$ Spatial Offset & (NB387$<24.5$ \& $I_{814}$$<26.5$) & 106 \\ 
    & (NB387$<24.5$ \& $H_{160}$$<26.5$) & 40 \\ \hline
    Ellipticity & ($I_{814}$$<25.0$ \& $R_e>0.\!\!^{\prime\prime}09$) & 41 \\
    & ($H_{160}$$<25.0$ \& $R_e>0.\!\!^{\prime\prime}18$) & 35 
\enddata
\tablecomments{Columns: (1) Quantity. (2) Magnitude and size cuts applied in each investigation. (3) Number of LAEs.} 
\label{table_sample}
\end{deluxetable*}

\begin{figure}[t!]
  \begin{center}
    \includegraphics[width=70mm]{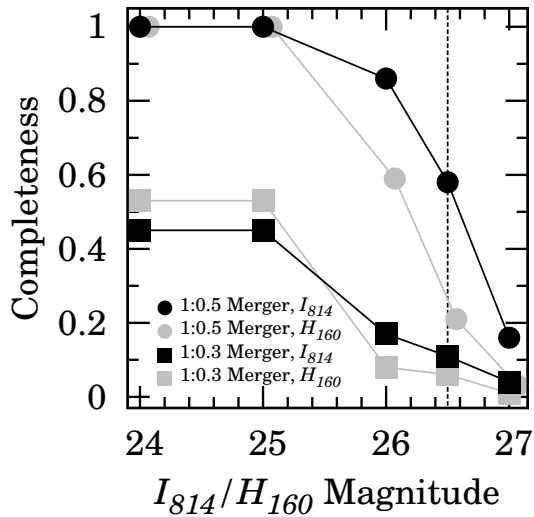}
  \end{center}
  \caption[EW and UV magnitude]{{\footnotesize Completeness of merger identification as a function of $I_{814}$ (black lines) and $H_{160}$ (gray lines) magnitudes. Filled circles and squares denote the completeness with component's flux ratios of $1$:$0.5$ and $1$:$0.3$, respectively. The dashed vertical line is a magnitude cut of $26.5$ mag for the close-pair method. The gray circles are slightly shifted along x-axis for the sake of clarity. }}
  \label{fig_merger_completeness}
\end{figure}

We carry out Monte Carlo simulations with artificial galaxy pairs to estimate detection completeness of major mergers. We consider two cases of flux ratios in major merger components, $0.3$--$1$ and $0.5$--$1$. We create 100 galaxy pairs with {\tt GALLIST} and {\tt MKOBJECTS} in {\tt IRAF} package in each bin of $I_{814}$/ $H_{160}$ magnitudes. In this procedure, we make these artificial galaxies at a redshift, and do not take into account the distance along the line-of-sight between components. This is because we aim to simply estimate the detection completeness of the fainter component as a function of {\it HST}-band magnitudes. On the other hand, our selection radius efficiently finds intrinsically-interacting objects, minimizing the chance projection rate of $\sim10$\% \citep[see ][]{2012ApJ...753...95B}. The created galaxy pairs are embedded into cutout images of randomly-selected blank fields. We detect these pairs in the same manner as for LAEs. 

The estimated completeness is shown in Figure \ref{fig_merger_completeness}. The artificial merger events with a flux ratio of $0.5$--$1$ are reproduced well for pairs brighter than $26.5$ mag in $I_{814}$ ($>50$\%). The completeness in $H_{160}$ is approximately half of that in $I_{814}$ at $26.0-26.5$ mag. This leads to the difference of derived merger fractions in $I_{814}$ and $H_{160}$ in \S \ref{subsec_result_merger}. In the case of the mergers with a flux ratio of $0.3$--$1$, the completeness is only $\sim50$\% even at $<24$ mag in $I_{814}$/$H_{160}$. This is because the magnitudes are severely underestimated for fainter components in the $0.3$:$1$ merger. 

As a result, we use $426$ and $237$ LAEs brighter than $26.5$ mag in $I_{814}$ and $H_{160}$, respectively, in the close-pair method. Table \ref{table_sample} summarizes the numbers of used LAEs including those in the following analyses. We provide the derived merger fraction and its dependence on Ly$\alpha$ EW in \S \ref{subsec_result_merger}.

\subsubsection{Morphological Index Method}\label{subsubsec_method_indices}

The non-parametric morphological indices have been widely utilized to characterize the structure and morphology of nearby and high-$z$ galaxies \citep[e.g., ][]{2007ApJS..172..270C,2007ApJS..172..468Z,2007ApJS..172..406S}. 

The $CAS$ system consists of the concentration ($C$), the asymmetry ($A$), and the smoothness ($S$) proposed by \citet{1996ApJS..107....1A,2000ApJ...529..886C,2003ApJS..147....1C}. Concentration $C$ is an index representing how much the flux concentrates into the galaxy's center. We calculate $C$ in the definition of \citet{2003ApJS..147....1C}, $C=5\log(r_{80}/r_{20})$, where $r_{80}$ and $r_{20}$ are the radii which contain $80$\% and $20$\% of the total flux of the galaxy, respectively. Asymmetry $A$ quantifies the degree of the rotational symmetry of the galaxy's light profile. It is calculated by 

\begin{equation}
A \equiv A_{\rm obj} - A_{\rm sky} = \frac{\Sigma | F - F_{180}|}{\Sigma |F|} - \frac{\Sigma | B - B_{180}|}{\Sigma |B|}, 
\end{equation}

where $F$ and $F_{180}$ ($B$ and $B_{180}$) are the original image of galaxy (sky background) and its image rotated by $180^{\circ}$ around the galaxy's center, respectively. The value of $A$ ranges from zero to one. Asymmetry becomes zero for a galaxy with a completely rotationally symmetric light profile. The first term in the definition of $A$ is the asymmetry of a galaxy.  The second term is the apparent asymmetry caused by sky background. We use average values of $A_{\rm sky}$  as representative $A_{\rm sky}$ in each field \citep[e.g., ][]{2007ApJS..172..406S}. Both of the $A_{\rm obj}$ and $A_{\rm sky}$ are computed by using all pixels contained within $1.5$ Petrosian radius of a targeted galaxy \citep{1976ApJ...209L...1P}. The rotational center is defined to be the position minimizing $A$ in the $3\times3$ grid searching method \citep{2000ApJ...529..886C}. The determined rotational center is also applied to the calculation of $C$. Because smoothness $S$ is not able to be correctly calculated for high-$z$ galaxies \citep{2004AJ....128..163L,2009MNRAS.397..208C}, we do not use $S$ in our analysis. 

To check the adequacy of our calculation, we compute $CA$ for other galaxies whose morphological indices have been already derived in previous studies. \citet{2007ApJS..172..270C} have calculated the indices for $\sim23000$ galaxies at low-$z$ in the COSMOS-Wide field. From their sample, we extract 300 galaxies whose $I_{814}$ magnitudes are comparable to those of our LAEs, and calculate their $CA$. Our calculation reproduces well the results of \citet{2007ApJS..172..270C}. We apply $1\sigma$ standard deviation from the $CA$ values obtained in \citet{2007ApJS..172..270C} to errors of the derived indices for our LAE sample. 

\begin{figure}[t!]
  \begin{center}
    \includegraphics[width=70mm]{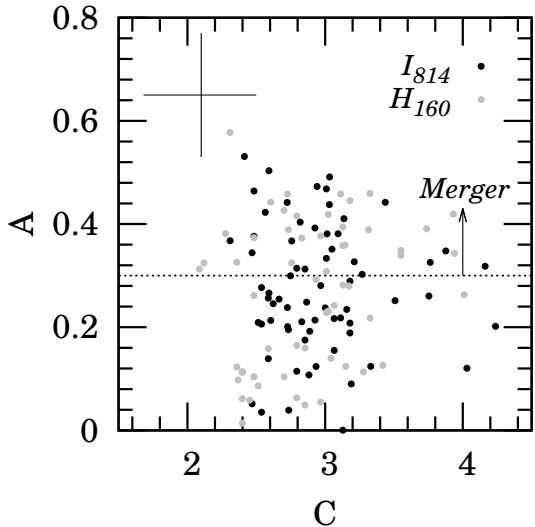}
  \end{center}
  \caption[C vs A]{{\footnotesize Concentration ($C$) and asymmetry ($A$) for counterparts in $I_{814}$ (black circles) and $H_{160}$ (gray circles). The region above the dashed line ($A=0.30$) indicates the merger regime. The error bar in the upper corner represents the typical uncertainty in individual objects. }}
  \label{fig_c_a}
\end{figure}

The morphological indices are not able to be robustly calculated for objects with a low $S/N$ \citep[e.g., ][]{2007ApJS..172..270C}. To obtain reliable values of the indices, we use LAEs with $I_{814}$ and $H_{160}$ magnitudes brighter than $25.0$ mag in our $CA$ calculation. The magnitude cut of $25.0$ mag has been usually applied in previous morphological studies with {\it HST} data \citep[e.g., ][]{2007ApJS..172..270C}. We also exclude objects whose half light radii in $I_{814}$ and $H_{160}$ are smaller than $0.\!\!^{\prime\prime}09$ and $0.\!\!^{\prime\prime}18$, which are unresolved with ACS and WFC3, respectively. We use 41 in $I_{814}$ and 35 LAEs in $H_{160}$, respectively, that meet all of these selection criteria. Representative examples of objects with a high and low $A$ value are shown in Figure \ref{fig_merger_example}. The calculated morphological indices of our LAE sample are shown in Figure \ref{fig_c_a}. The distributions of indices in these parameter spaces are quite similar to results of Lyman Break Galaxies (LBGs) at $z\sim2-3$ \citep[e.g., ][]{2012ApJ...745...85L}. 

In order to classify mergers, we adopt the following criterion, 

\begin{equation}\label{eq_criterion_asymmetry}
A > 0.30, 
\end{equation}

which is defined by \citet{2008MNRAS.391.1137L} for high-$z$ galaxies. We do not use the $C$ parameter for the merger classification, but we make the $C$--$A$ diagram, simply for the sake of clarity (Fig. \ref{fig_c_a}). 

In the $A$ classification, we identify mergers which are brighter than a specific {\tt MAG\_APER} magnitude ({\tt MAG\_APER} cut) in addition to the {\tt MAG\_AUTO} cut. The {\tt MAG\_AUTO} cut mistakenly selects objects with extremely-low surface brightness. The asymmetry parameter might be overestimated for these objects due to their diffuse structure. We also derive the merger fraction for the sample selected in the {\tt MAG\_APER} cut to eliminate the diffuse objects. We use a $0.\!\!^{\prime\prime}3$ aperture to calculate {\tt MAG\_APER}. The number of selected counterparts significantly decreases, especially in the $H_{160}$ data. Even in this case, we find the similar trend of the Ly$\alpha$ EW dependence as in the {\tt MAG\_AUTO} cut. 

We discuss the merger fraction based on the index classification in \S \ref{subsec_result_merger}.

\begin{figure}[t!]
  \begin{center}
    \includegraphics[width=60mm]{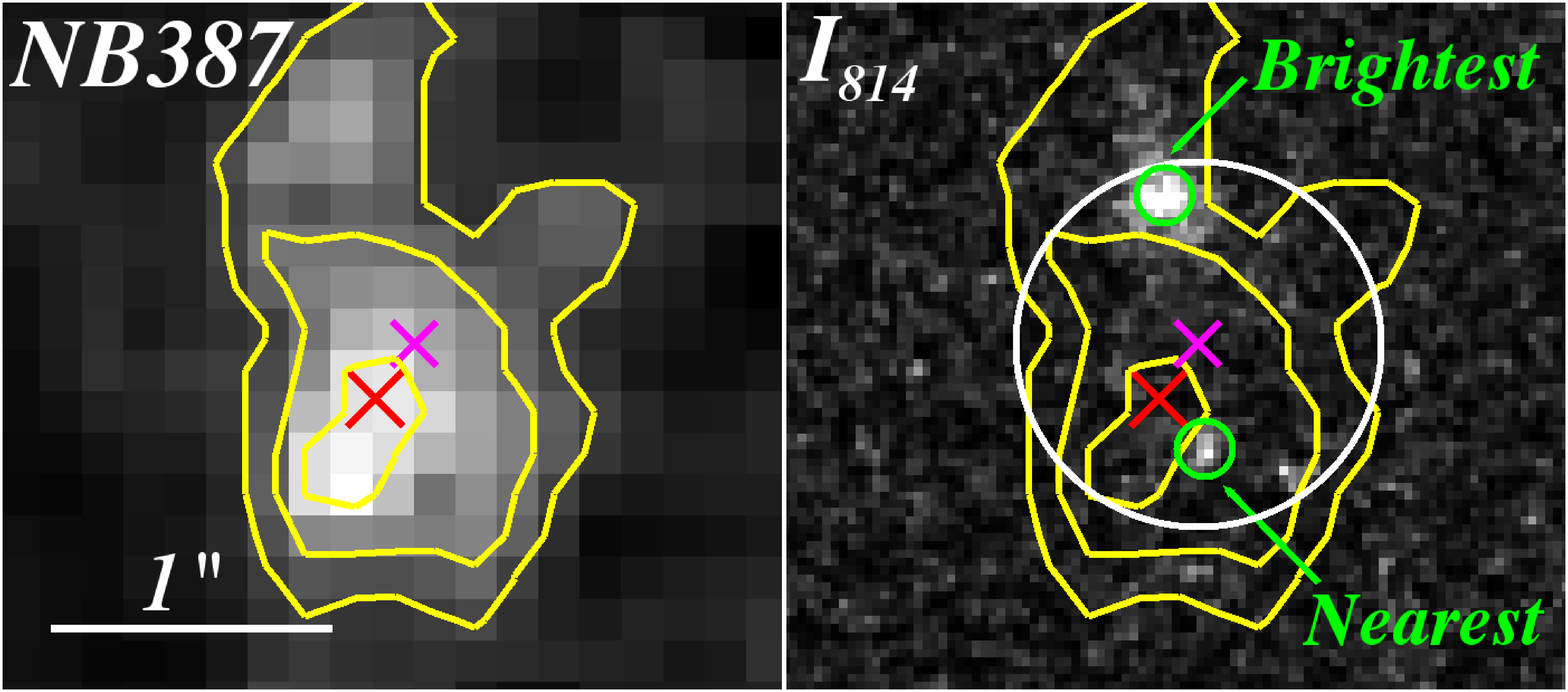}\\
    \includegraphics[width=60mm]{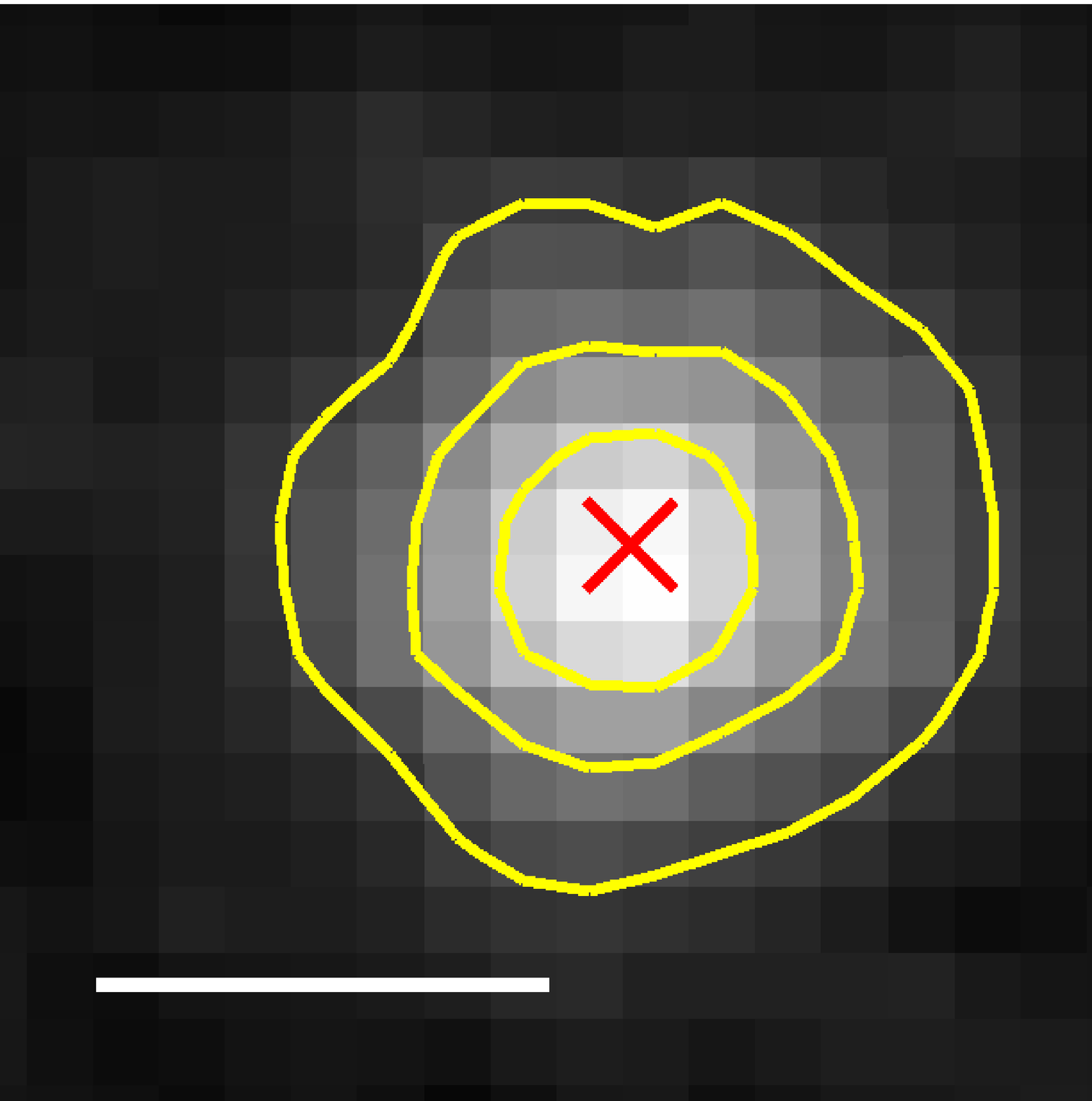}\\
    \includegraphics[width=60mm]{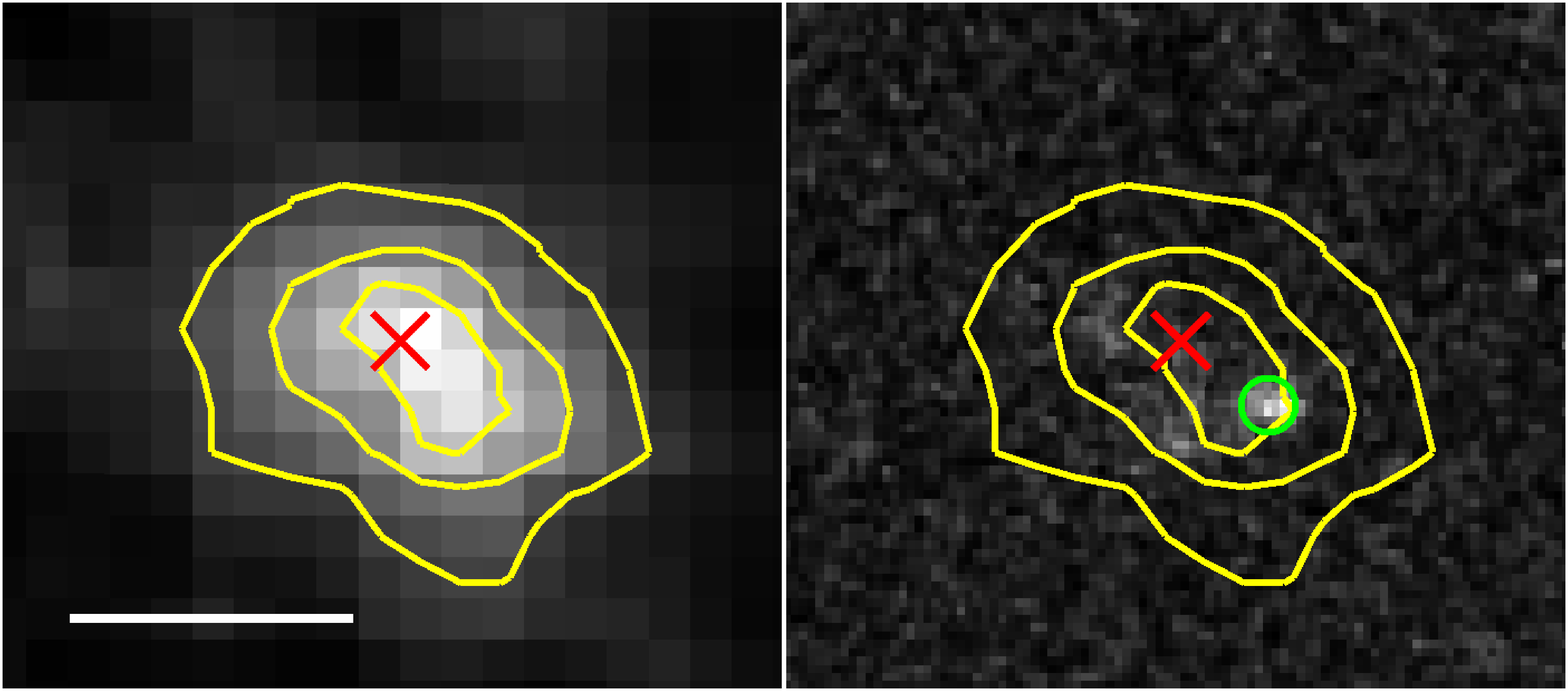}
  \end{center}
    \caption[NB contour]{{\footnotesize NB387 (left) and $I_{814}$ (right) images of three example LAEs. The yellow contours indicate the isophotal area in the NB387 images. Green open circles denote the continuum counterparts in each $I_{814}$ image. ({\it Top}) An LAE whose NB centroid is redefined. Magenta crosses depict the original position determined in the NB387 imaging studies \citep{2012ApJ...745...12N, 2013ApJ...769....3N}. Red crosses represent the central position redefined from the {\tt SExtractor} detection with a higher {\tt DETECT\_THRESH} value in the NB387 images. The NB centroid shifts by $\sim0.\!\!^{\prime\prime}2$ towards the peak of light profile. See \S \ref{subsec_method_offset}. The top and bottom green circles are the {\it brightest} and {\it nearest} continuum counterparts, respectively. The spatial offsets from the redefined NB center are $\sim0.\!\!^{\prime\prime}7$ for the brightest and $\sim0.\!\!^{\prime\prime}2$ for the nearest counterpart. The white circle indicates our selection radius of $0.\!\!^{\prime\prime}65$. ({\it Middle}) An LAE whose Ly$\alpha$ centroids are aligned well on the position of continuum emission. ({\it Bottom}) An LAE having a large $\delta_{\rm Ly\alpha}$ Ly$\alpha$ spatial offset. The Ly$\alpha$ EW (the Ly$\alpha$ spatial offset) of the object in the middle and bottom panels are $218$ ($0.\!\!^{\prime\prime}11$) and $38$\,\AA\, ($0.\!\!^{\prime\prime}40$), respectively. White bars in the lower corner of each NB387 image indicate 1$^{\prime\prime}$, corresponding to $\sim8.3$ kpc at $z=2.2$. North is up and east is to the left. }}
  \label{fig_contour}
\end{figure}

\subsection{Ly$\alpha$ Spatial Offset Between Rest-Frame UV/Optical Continuum and Ly$\alpha$ Emission}\label{subsec_method_offset}

We calculate projected distance between the rest-frame UV/optical continuum positions and the centroids of Ly$\alpha$ emission. In this calculation, we consider two types of components in stellar-continuum emission. One is the {\it brightest} components in all sources within the selection radius, and the other is the {\it nearest} ones from Ly$\alpha$ centroids among objects identified in our detection criteria of {\tt SExtractor} (the top panel of Figure \ref{fig_contour}). The central position of each {\it HST} cutout image corresponds to Ly$\alpha$ centroids. We calculate $\delta_{\rm Ly\alpha}$ from the coordinates of sources in the cutout images. 

We find that the Ly$\alpha$ centroids are slightly shifted in a direction toward extended and diffuse light structure in several NB images, as shown in the top panel of Figure \ref{fig_contour}. The central position of LAEs has been determined by performing the source detection with {\tt SExtractor} in entire $\sim30^{\prime}\times30^{\prime}$ Suprime-Cam images \citep{2012ApJ...745...12N} (the magenta cross). The small positional offsets of Ly$\alpha$ are certainly caused by setting {\tt DETECT\_THRESH} to a relatively-low value ($2.0\sigma$) for the LAE selection in the wide images. They have used a typical value of {\tt DETECT\_THRESH} in selections for high-$z$ galaxies. That value is too low to estimate the peak position of Ly$\alpha$.
 
\begin{figure}[t!]
  \begin{center}
    \includegraphics[width=80mm]{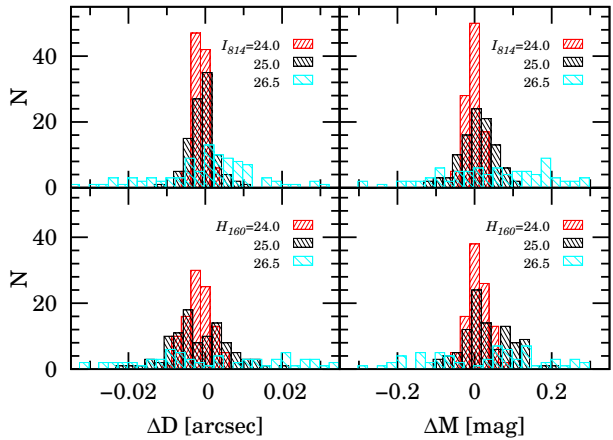}
  \end{center}
  \caption[EW and UV magnitude]{{\footnotesize Difference between input and output values of positions (left) and magnitudes (right) in $I_{814}$ (top panels) and $H_{160}$ data (bottom panels) for artificial galaxies. A hundred of artificial galaxies are created in each magnitude of $I_{814}, H_{160}=24$ (red), $25$ (black), and $26.5$ (cyan). The histograms are slightly shifted along x-axis for the sake of clarity. }}
  \label{fig_sim_offset_hst}
\end{figure}

In order to estimate the peak position of Ly$\alpha$, we carry out the source detection with a higher {\tt DETECT\_THRESH} value ($2.5\sigma$) in each NB cutout image than the value used in the NB selection. This procedure is very efficient in calculating the Ly$\alpha$ spatial offset from the position where Ly$\alpha$ is emitted most efficiently. This position corresponds to the location where the galaxy is brightest in Ly$\alpha$. In the re-detection process, NB centroids are slightly shifted for several objects, and we obtain redefined values of $\delta_{\rm Ly\alpha}$. We adopt the original centroids for objects with the Ly$\alpha$ positional difference smaller than $0.\!\!^{\prime\prime}1$ corresponding to $\sim0.5$ pixel in NB images.

\begin{figure}[t!]
  \begin{center}
    \includegraphics[width=75mm]{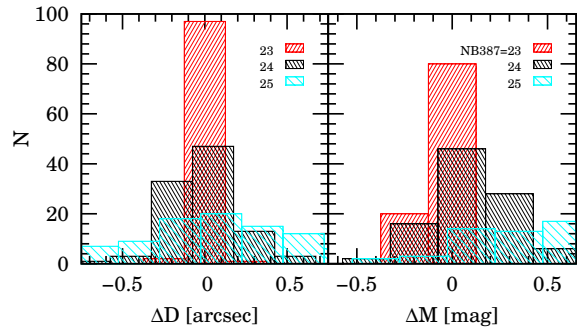}
  \end{center}
  \caption[EW and UV magnitude]{{\footnotesize Same as Figure \ref{fig_sim_offset_hst}, but for Subaru/NB387 images. Artificial galaxies are created in each magnitude of NB387$=23$, $24$, and $25.5$.}}
  \label{fig_sim_offset_nb}
\end{figure}

We create artificial galaxies to estimate the positional errors using the same method as described in \S \ref{subsubsec_method_closepair}. Figures \ref{fig_sim_offset_hst} and \ref{fig_sim_offset_nb} show the estimated positional errors of the $I_{814}$/$H_{160}$ and NB387 images, respectively. These figures indicate that the positional uncertainties tend to be larger for fainter objects. In the {\it HST} images, the positional error is less than $\sim\pm0.\!\!^{\prime\prime}02$ at $I_{814}$/$H_{160}$ magnitudes brighter than $26.5$. In contrast, the NB387 images have a large positional error of $\sim0.\!\!^{\prime\prime}3$ (at $1\sigma$) even at NB387$=24.5$. This large uncertainty is due to the relatively-large seeing sizes ($\sim0.\!\!^{\prime\prime}8$) in the NB data taken by the ground-based observations. 

To obtain reliable Ly$\alpha$ spatial offsets, we use objects with $I_{814}$/$H_{160}$$<26.5$ and NB387$<24.5$ in this analysis. In this case, 106 and 40 LAEs are selected in $I_{814}$ and $H_{160}$, respectively. Moreover, no systematic error in the simulation ensures a statistical investigation of the Ly$\alpha$ spatial offset. NB387 and $I_{814}$ images of example galaxies are shown in Figure \ref{fig_contour}. We provide the dependence of $\delta_{\rm Ly\alpha}$ on Ly$\alpha$ EW in \S \ref{subsec_result_offset}.

\begin{figure}[t!]
  \begin{center}
    \includegraphics[width=60mm]{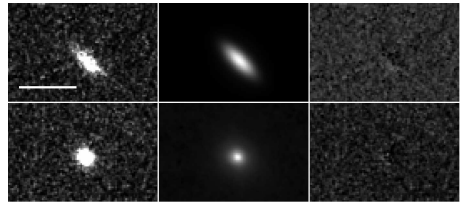}
  \end{center}
  \caption[NB contour]{{\footnotesize Representative examples with large and small ellipticity in {\tt GALFIT} fitting. The left, center, and right panels represent $I_{814}$ images, the best-fit S\'ersic profiles, and their residual images, respectively. White bars in the lower corner of each NB387 image indicates 1$^{\prime\prime}$.}}
  \label{fig_galfit}
\end{figure}

\begin{figure*}[t!]
  \begin{center}
    \includegraphics[width=110mm]{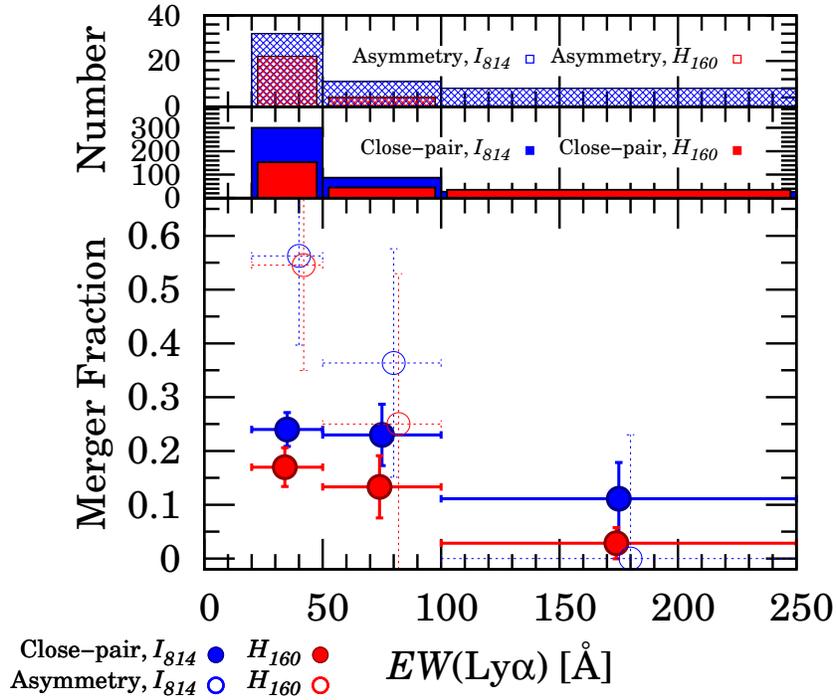}
  \end{center}
  \caption[Merger fraction and Ly$\alpha$ equivalent width. ]{{\footnotesize Merger fraction of $z=2.2$ LAEs as a function of Ly$\alpha$ EW. Blue and red symbols indicate the merger fractions in $I_{814}$ and $H_{160}$, respectively. Filled and open circles denote the merger fraction in the close-pair method and the $A$ classification, respectively.  Error bars in each plot include the Poisson statistical error. Error bars in Ly$\alpha$ EW indicate the bin widths. The symbols are slightly shifted along x-axis for the sake of clarity. There are no LAEs in $H_{160}$ in the highest EW bin for the $A$ classification, because no mergers are found in this EW bin. The histograms in the upper and middle panels show the number of LAEs in each Ly$\alpha$ EW bin in the $A$ classification and the close-pair method, respectively. }}
  \label{fig_ew_mergerrate}
\end{figure*}

\subsection{Ellipticity}\label{subsec_method_ellipticity}

We measure the ellipticity $\epsilon$ of the counterparts in $I_{814}$/$H_{160}$ using the {\tt GALFIT} software \citep{2002AJ....124..266P,2010AJ....139.2097P}. The ellipticity is defined as $\epsilon= 1-b/a$, where $a$ and $b$ are the major and minor axes, respectively. The ellipticity is calculated for both of the brightest and nearest components in the same manner as in \S \ref{subsec_method_offset}. In this calculation, we use objects whose $I_{814}$/$H_{160}$ magnitudes are brighter than $25.0$ and half light radii are larger than the typical PSF sizes of each band. 

\begin{figure}[t!]
  \begin{center}
    \includegraphics[width=70mm]{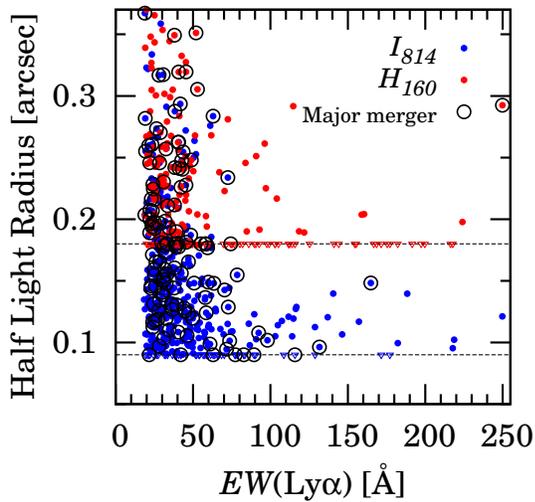}
  \end{center}
  \caption[EW and UV magnitude]{{\footnotesize Half light radius and Ly$\alpha$ EW of continuum counterparts in $I_{814}$ (blue) and $H_{160}$ (red) data. Black open circles mark objects with multiple components in a selection radius of $0.\!\!^{\prime\prime}65$ (major mergers). Horizontal lines indicate typical PSF sizes of $I_{814}$ and $H_{160}$ bands. Objects smaller than the typical PSF sizes are placed at $0.\!\!^{\prime\prime}09$ in $I_{814}$ and $0.\!\!^{\prime\prime}18$ in $H_{160}$ (open inverted triangles). }}
  \label{fig_hlr_ew}
\end{figure}

The profile fitting are performed in the similar manner as \citet{2011ApJ...743....9G}. The counterparts are fitted to a S\'ersic profile. Some initial parameters are needed in the {\tt GALFIT} fitting. The coordinates $(x_c, y_c)$, total magnitude $m$, axis ratio $q(=b/a)$, position angle $PA$, and half light radius $r_e$ of each counterpart are input into the {\tt GALFIT} configuration file as initial parameters. These initial parameters are estimated with {\tt SExtractor} prior to the {\tt GALFIT} fitting. The S\'ersic index is set to $n=4$ (i.e., de Vaucouleurs profile) as an initial value, while initial S\'ersic index does not affect the fitting results \citep{2011ApJ...736...92Y,2012ApJ...761...19Y}. In the fitting procedure, we also allow the following parameters to move in the ranges, $24<m<29$, $0.1<r_e<15$ pixels, $0.1<n<15$, $0.1<q<1$, $\Delta x<2$, and $\Delta y<2$. We create PSF images for $I_{814}$ and $H_{160}$ data by stacking 100 bright isolated point sources. {\tt GALFIT} outputs best-fit parameters corrected for PSF broadening. Figure \ref{fig_galfit} shows examples of original $I_{814}$ images, the best-fit S\'ersic profiles, and their residual images. As shown in Figure \ref{fig_galfit}, the {\tt GALFIT} fits well the stellar-continuum emission to S\'ersic profiles. 

We provide the dependence of ellipticity on Ly$\alpha$ EW in \S \ref{subsec_result_ellipticity}.

\section{RESULTS}\label{sec_results}

In this section, we show the derived merger fraction, Ly$\alpha$ spatial offset, and ellipticity as a function of $EW{\rm (Ly\alpha)}$. We divide our LAE sample into three $EW{\rm (Ly\alpha)}$ bins, $20-50$, $50-100$, and $>100$ \AA.

\subsection{Dependence of Merger Fraction on Ly$\alpha$ EW}\label{subsec_result_merger}

We first compare the merger fractions of the entire sample with results from previous studies for LAEs at $z\sim2-6$. The merger fractions in $I_{814}$ and $H_{160}$ are $0.23\pm0.02$ and $0.14\pm 0.02$, respectively, in the close-pair method. These values are broadly consistent with other studies in the similar methods \citep[e.g., ][]{2009ApJ...701..915T,2007ApJ...667...49P,2010ApJ...711..928C,2009ApJ...705..639B}. \citet{2007ApJ...667...49P} investigated rest-frame UV morphologies of nine LAEs at $4\lesssim z\lesssim5.7$ based on the $C-A$ classification. They find that $\sim30-40$\% of the sample show clumpy, complex, or morphologically-disturbed structures. \citet{2009ApJ...701..915T} present that only two out of $\sim50$ LAEs at $z=5.7$ have double-component structures, and $\sim50$\% of the sample are spatially extended in the rest-frame UV. \citet{2009ApJ...705..639B} obtain that at least $\sim17$\% of their 120 LAEs at $z=3.1$ have multiple components. \citet{2010ApJ...711..928C} claim that $>30$\% of $z\sim0.3$ LAEs show merger features. Theoretically, the dark matter simulation combined with a physical model of \citet{2011MNRAS.418.2196T} predicts that the merger fraction of $z\sim3$ LAEs is $\sim0.20$ after matching the merger mass ratio with that of observational studies. Our merger fractions are also nearly the same as those of LBGs at $1.5\lesssim z\lesssim3$ estimated with the close-pair method  \citep[$\sim 0.05-0.2$; ][]{2012ApJ...745...85L}. Our merger fractions in the $A$ classification are also similar to that of their LBG sample estimated with the same method. 

Next, we examine the dependence of the merger fraction on Ly$\alpha$ EW. Figure \ref{fig_ew_mergerrate} shows the derived merger fractions in each $EW{\rm (Ly\alpha)}$ bin. Error bars in each plot include the Poisson statistical errors. We find that the merger fraction does {\it not} significantly increase with Ly$\alpha$ EW in all cases. Instead, merger fractions decrease from the lowest to the highest EW bin over the 1$\sigma$ error bars in several cases. The merger fractions in $H_{160}$ are a factor of $\sim2-3$ lower than those in $I_{814}$ in the corresponding methods. This is likely to be caused by the difference of completeness (\S \ref{subsubsec_method_closepair}). In addition to the completeness effect, the difference of the merger fractions could be explained by the shapes of SEDs, since the $I_{814}$- and the $H_{160}$-band data trace the rest-frame UV and optical stellar continuum emission, respectively. We also examine the total merger fraction derived in combination with the close-pair and morphological index methods. We define the total merger fraction as the logical sum of mergers identified by the close-pair method and $A$ classification. However, we do not find the increase of merger fraction with $EW{\rm (Ly\alpha)}$, similar to Figure \ref{fig_ew_mergerrate}. We additionally calculate the Ly$\alpha$ EW from the total magnitudes in the NB data, and evaluate an effect of the Ly$\alpha$ flux loss in the aperture photometry on the Ly$\alpha$ dependence. We compute the total magnitudes with {\tt MAG\_AUTO} of {\tt SExtractor}. Even in this test, we still do not find a trend that the merger fraction increases with Ly$\alpha$ EW. 

We change the selection radius to $r_{\rm sel} = 1.\!\!^{\prime\prime}5$ ($\sim13$ kpc at $z=2.2$), in order to check possible differences in merger fraction between searching radii. The $1.\!\!^{\prime\prime}5$ radius is the same as that used in \citet{2012ApJ...745...85L}. However, we do not find a significant rise of merger fraction even in the larger radius. The merger fractions with $r_{\rm sel}=1.\!\!^{\prime\prime}5$ are a factor of $\sim2$ higher than those with  $r_{\rm sel}=0.\!\!^{\prime\prime}65$. Adopting the $1.\!\!^{\prime\prime}5$ aperture, the merger fractions increase to $0.35\pm0.03$ in $I_{814}$ and $0.30\pm0.03$ in $H_{160}$. We confirm that the trend of Figure \ref{fig_ew_mergerrate} is similarly found in the results of $r_{\rm sel}=1.\!\!^{\prime\prime}5$ aperture.

The dependence of the merger fraction on Ly$\alpha$ EW is also clearly shown in Figure \ref{fig_hlr_ew}. The figure illustrates a trend that objects with a larger Ly$\alpha$ EW have a smaller half light radius, as claimed by e.g., \citet{2012ApJ...759...29L} and \citet{2010A&A...514A..64P}, which is also justified by our statistical tests. The individual LAEs identified in the close-pair method are marked by the black circles. Figure \ref{fig_hlr_ew} clearly exhibits the small number of mergers at $EW{\rm (Ly\alpha)}>100$ \AA. Note that the decline in the merger fraction in high $EW{\rm (Ly\alpha)}$ bins is caused by many incomplete detections of fainter merger components in LAEs with a high EW. Our magnitude cut of $26.5$ mag ensures no bias in $I_{814}$/$H_{160}$ magnitudes between EW bins (Fig. \ref{fig_ew_mag}). The merger completeness is considered to be almost constant in all of the EW bins.

\begin{figure*}[t!]
  \begin{center}
    \includegraphics[width=115mm]{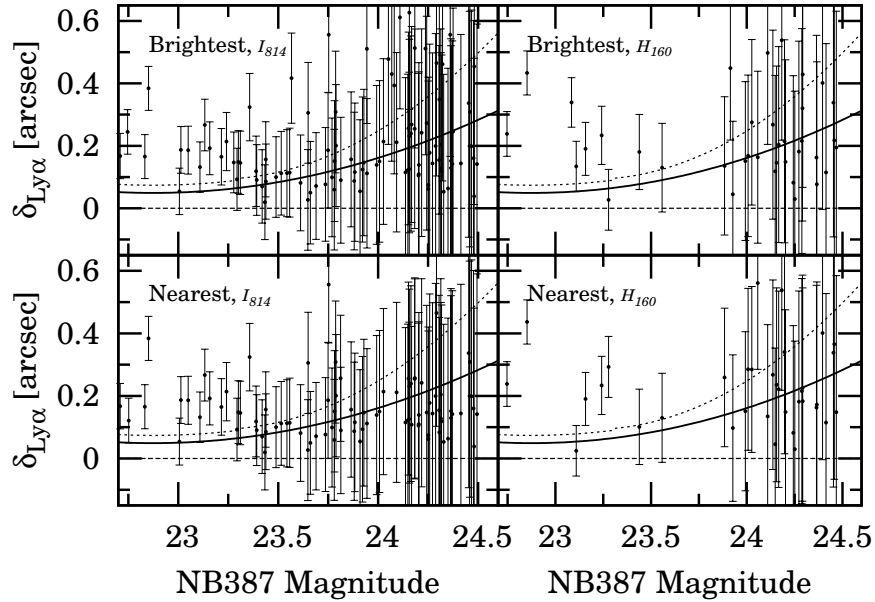}
  \end{center}
  \caption[Spatial offset of rest-UV/optical continuum emission from Ly$\alpha$ centroids.]{{\footnotesize Spatial offset  as a function of NB387 magnitude for the brightest counterparts in $I_{814}$ (upper-left), and $H_{160}$ (upper-right), and the nearest ones in $I_{814}$ (bottom-left), and $H_{160}$ (bottom-right). The bold and dashed curves denote mean values of $\delta_{\rm Ly\alpha}$, and their $1\sigma$ standard deviations, respectively, estimated from the Monte Carlo simulations in \S \ref{subsubsec_method_closepair}. The error bars in $\delta_{\rm Ly\alpha}$ are based on the positional uncertainties quantified in these simulations. }}
  \label{fig_offset_mag_tile}
\end{figure*}

\subsection{Dependence of Ly$\alpha$ Spatial Offset on Ly$\alpha$ EW}\label{subsec_result_offset}

We investigate $\delta_{\rm Ly\alpha}$ of Ly$\alpha$ spatial offset and examine whether $\delta_{\rm Ly\alpha}$ is produced by measurement errors or a real signal. Figure \ref{fig_offset_mag_tile} shows $\delta_{\rm Ly\alpha}$ as a function of NB387 magnitude. We include statistical errors estimated from the Monte Calro simulation (\S \ref{subsec_method_offset}) in $\delta_{\rm Ly\alpha}$. We find a tendency that the error in $\delta_{\rm Ly\alpha}$ becomes larger for the objects with a fainter NB387 magnitude, but successfully identify that several LAEs have an offset beyond our statistical errors for relatively-bright objects. The identification of the large $\delta_{\rm Ly\alpha}$ objects could not be due to large scatters in $\delta_{\rm Ly\alpha}$, which is justified by our non-parametric Kolmogorov-Smirnov (KS) tests between the LAEs and artificial galaxies in each NB387 magnitude bin. The KS probabilities are calculated to be $P_{\rm KS} \lesssim 0.05$. 

\begin{figure}[t!]
  \begin{center}
    \includegraphics[width=80mm]{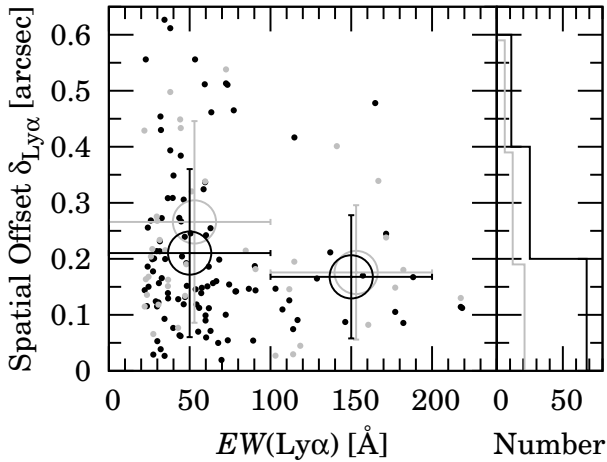}
  \end{center}
  \caption[Spatial offset of rest-UV/optical continuum emission from Ly$\alpha$ centroids.]{{\footnotesize Spatial offset between the rest-frame UV/optical continuum emission of the brightest counterparts and their Ly$\alpha$ centroids. Black and gray filled circles indicate counterparts in the $I_{814}$ and $H_{160}$ cutout images, respectively. Large black and gray circles are the average values of $\delta_{\rm Ly\alpha}$ in subsamples of $EW<100$,  and $>100$\,\AA. The right panel shows histograms for the number of LAEs. The histograms are slightly shifted along y-axis for the sake of clarity. The position of Ly$\alpha$ emission is redefined in the {\tt SExtractor} detection with a higher {\tt DETECT\_THRESH} value. The detail is described in \S \ref{subsec_method_offset}.}}
  \label{fig_offset_ew_redefined}
\end{figure}

\begin{figure}[t!]
  \begin{center}
    \includegraphics[width=80mm]{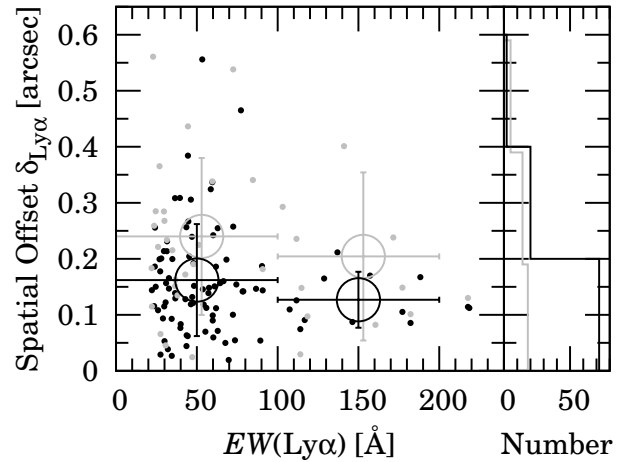}
  \end{center}
  \caption[Spatial offset of rest-UV/optical continuum emission from Ly$\alpha$ centroids (redefined NB center).]{{\footnotesize Same as Figure  \ref{fig_offset_ew_redefined}, but for the nearest counterparts.}}
  \label{fig_offset_ew_nearest_redefined}
\end{figure}

Next, we investigate the dependence of $\delta_{\rm Ly\alpha}$ on Ly$\alpha$ EW for the brightest continuum sources in Figure \ref{fig_offset_ew_redefined} and the nearest ones in Figure \ref{fig_offset_ew_nearest_redefined}. We find that there are few LAEs with a high EW and a large $\delta_{\rm Ly\alpha}$. LAEs with a high Ly$\alpha$ EW tend to have a single continuum counterpart, as described in \S \ref{subsec_result_merger}. The distribution of $\delta_{\rm Ly\alpha}$ for high EW objects does not depend strongly on whether we use the brightest or the nearest continuum counterparts. 

We carry out the KS test in order to evaluate whether $\delta_{\rm Ly\alpha}$ is statistically related to Ly$\alpha$ EW. We calculate the KS probability that LAEs with Ly$\alpha$ EW $>100$\,\AA\, and $<100$\,\AA\, are drawn from the statistically-same distribution of the Ly$\alpha$ spatial offset. We summarize the KS probabilities, $P_{\rm KS}$, in Table \ref{table_kstest}. The $P_{\rm KS}$ values are $~0.05-0.1$ in the case of the original Ly$\alpha$ centroid. In the $I_{814}$ data, the low $P_{\rm KS}$ values indicate that the two groups of $\delta_{\rm Ly\alpha}$ are drawn from statistically-different distribution. Even in the case of the redefined values of $\delta_{\rm Ly\alpha}$, the probabilities are not significantly changed ($P_{\rm KS}\sim0.1-0.3$).

\begin{deluxetable}{ccc}
\tabletypesize{\scriptsize}
\tablecaption{Results of KS Test}
\tablehead{  \colhead{Quantity} & \colhead{Counterpart and Band} & \colhead{$P_{KS}$} \\ 
\colhead{(1)}& \colhead{(2)}& \colhead{(3)}} 

\startdata
    Ly$\alpha$ Spatial Offset & Brightest ($I_{814}$) & 0.262 (0.096) \\ 
    & Brightest ($H_{160}$) & 0.285 (0.097) \\ 
    & Nearest ($I_{814}$) & 0.177 (0.145) \\ 
    & Nearest ($H_{160}$) & 0.147 (0.050) \\ \hline
    Ellipticity & Brightest ($I_{814}$) & 0.579 \\
    & Brightest ($H_{160}$) & -\tablenotemark{a} \\ 
    & Nearest ($I_{814}$) & 0.564 \\ 
    & Nearest ($H_{160}$) & -\tablenotemark{a} 
\enddata
\tablecomments{Columns: (1) Quantity. (2) Type of continuum counterparts ({\it brightest} or {\it nearest}), and used {\it HST} band. (3) KS probability that LAEs with Ly$\alpha$ EW $<100$\,\AA\, and $>100$\,\AA\, are drawn from the statistically-same distribution. The values in parentheses represent the probabilities for the distribution of $\delta_{\rm Ly\alpha}$ after correcting for the NB centroids with a higher {\tt DETECT\_THRESH}. See \S \ref{subsec_method_offset}. }
\tablenotetext{a}{KS probabilities cannot be calculated because only one object with $EW(Ly{\rm \alpha})>100$\,\AA\,matches the selection criteria in the $H_{160}$ data. }
\label{table_kstest}
\end{deluxetable}

\subsection{Dependence of Ellipticity on Ly$\alpha$ EW}\label{subsec_result_ellipticity}

\begin{figure}[t!]
  \begin{center}
    \includegraphics[width=70mm,angle=-90]{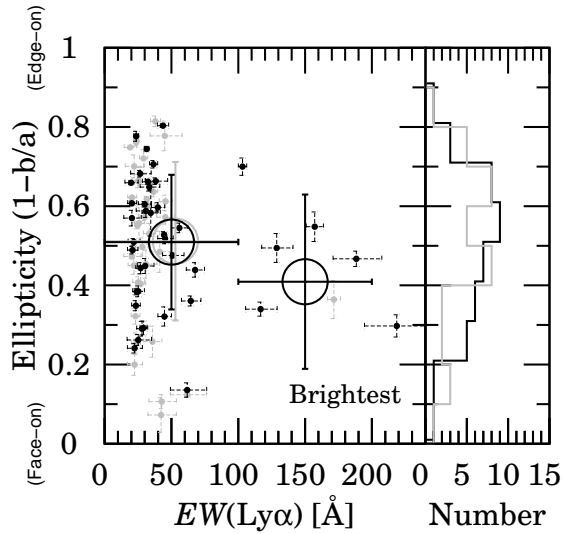}
  \end{center}
  \caption[Ellipticity and Ly$\alpha$ equivalent width. ]{{\footnotesize Ellipticity of the {\it brightest} continuum counterparts and its dependence on their Ly$\alpha$ EW. Black and gray filled circles indicate counterparts in $I_{814}$ and $H_{160}$ cutout images, respectively. Large black and gray circles are the average ellipticity in subsamples of $EW<100$,  and $>100$\,\AA. The right panel shows histograms for the number of LAEs. The histograms are slightly shifted along y-axis for the sake of clarity. }}
  \label{fig_ew_ellipticity_brightest}
\end{figure}

We show the ellipticity of the brightest and nearest continuum objects in Figures \ref{fig_ew_ellipticity_brightest} and \ref{fig_ew_ellipticity_nearest}, respectively. The average ellipticity of LAEs with $EW({\rm Ly\alpha})>100$\,\AA\, is smaller than that of objects with $EW({\rm Ly\alpha})<100$\,\AA. There is a possible trend that the LAEs with a large Ly$\alpha$ EW have a small ellipticity for both of the brightest and nearest components. The right panel of each figure displays histograms of the ellipticity. The ellipticity distribution is quite similar to that estimated by \citet{2011ApJ...743....9G} who have studied morphologies of LAEs at the similar redshift. 

We calculate the KS probability for the ellipticity in the similar manner as for the Ly$\alpha$ spatial offset in \S \ref{subsec_result_offset}. The probabilities are listed in Table \ref{table_kstest}.  The probabilities are calculated to be $0.5-0.6$ in the $I_{814}$ data. The high values indicate that LAEs with Ly$\alpha$ EW $>100$\,\AA\, and $<100$\,\AA\, have statistically-indistinguishable distributions. These probabilities suggest that the dependence of ellipticity on Ly$\alpha$ EW cannot be concluded in a statistical sense. The small sample size of LAEs with $EW({\rm Ly\alpha})>100$\,\AA\, may not allow us to obtain accurate KS probabilities.

\section{DISCUSSION} \label{sec_discussion}

\subsection{Ly$\alpha$ Enhancement by Major Merger}\label{subsec_discuss_merger_fraction}

In \S \ref{subsec_result_merger}, we find that the merger fraction of LAEs does not significantly increase with their Ly$\alpha$ EW. Instead, Figure \ref{fig_ew_mergerrate} shows the merger fraction decreases from $EW({\rm Ly\alpha})=20-100$ to $>100$\,\AA. However, our statistical analysis indicates that mergers are rare in the subsample of LAEs with a Ly$\alpha$ EW larger than $100$ \,\AA. Our result would suggest that the galaxy merger does not heavily affect the distribution of H {\sc i} gas and dust. On the contrary, the H {\sc i} clouds disturbed by a merger would envelop a central ionizing source instead of making holes in the gas shell. The nearly-uniform clouds might prevent Ly$\alpha$ photons from easily escaping from a galaxy. 

This is opposite to a trend that Ly$\alpha$ emission is enhanced by a galaxy merger. Several observational studies have examined a relationship between Ly$\alpha$ EW and galaxy merger in the LAE population.  For example, \citet{2010MNRAS.403.1020C}, and \citet{2013ApJ...775...99C} claim that Ly$\alpha$ emission is enhanced by galaxy merger. This trend is commonly based on the sense that a galaxy interaction triggers star formation, and disperses obscuring gas and dust in the system \citep[e.g., ][]{2013ApJ...775...99C}. 

\citet{2010MNRAS.403.1020C} carry out spectroscopic observations for 140 LBGs at $z\sim3$, and find serendipitously five LBG pairs with projected proper separations less than $15$ kpc. They additionally discover two LAEs with a close LBG in their MOS slitlets. One of these LAEs has a Ly$\alpha$ EW of 48 \,\AA. The separation between the LAE and its LBG companion is $22.7$ kpc. Another LAE has a Ly$\alpha$ EW of 140 \,\AA, but its LBG companion is not definitively confirmed by spectroscopy. This merger candidate has a relatively large projected separation of $40.1$ kpc between its components. In this survey, only one object with such a high EW has been found in the seven serendipitously-discovered close pairs ($\sim14$\%), if the LAE with $EW({\rm Ly\alpha})=140$ \,\AA\, is a genuine merger. 

Recently, \citet{2013ApJ...775...99C} have investigated three LAEs in the HETDEX sample. The Ly$\alpha$ EW of all the three LAEs exceeds $100$ \,\AA\, ($114\pm13$, $140\pm43$, and $206\pm65$ \,\AA) due to a unique LAE selection method of the HETDEX survey \citep{2011ApJS..192....5A}. In the {\it HST} images, two LAEs with $EW({\rm Ly\alpha})=114$ and $140$ \,\AA\, show close components with projected separations of $\sim5$ and $8$ kpc, respectively. One of these close components has also been spectroscopically confirmed to be at the same redshift as its central LAE. The LAE with the highest EW of $206$ \,\AA\, does not have a companion within $1^{\prime\prime}$ ($\sim8.2$ kpc)\footnote{A nearby continuum source is shown at a projected distance of $4^{\prime\prime}$ ($\sim33$ kpc) from the LAE.}. Thus, the merger fraction at $EW({\rm Ly\alpha})>100$ \,\AA\, is $0.67^{+0.33}_{-0.43}$ in their HETDEX sample. The error is based on the small number statistics \citep{1986ApJ...303..336G}. Our merger fraction at $EW({\rm Ly\alpha})>100$ is $0.23\pm0.08$ in $I_{814}$ in the similar searching radius of $1.\!\!^{\prime\prime}5$ ($\sim13$ kpc) to that of \citet{2013ApJ...775...99C}. The merger fractions are consistent within $1\sigma$ uncertainties. In the study of \citet{2013ApJ...775...99C}, the merger fraction at $EW({\rm Ly\alpha})>100$ \,\AA\, becomes probably higher due to the small number of sample objects. 

\begin{figure}[t!]
  \begin{center}
    \includegraphics[width=70mm,angle=-90]{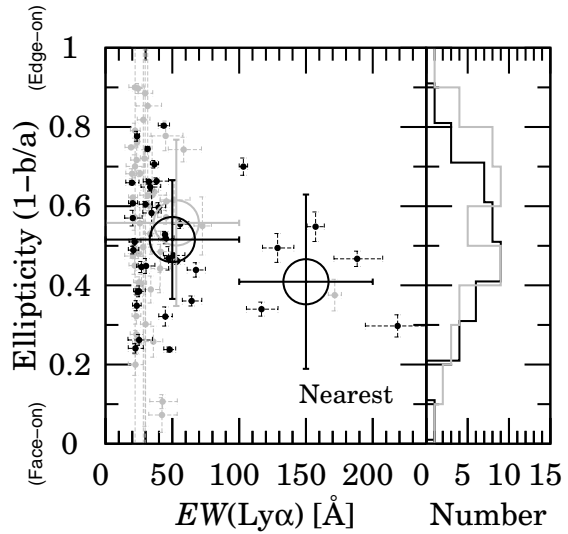}
  \end{center}
  \caption[Ellipticity and Ly$\alpha$ equivalent width. ]{{\footnotesize Same as Figure \ref{fig_ew_ellipticity_brightest}, but for the {\it nearest} continuum counterparts.}}
  \label{fig_ew_ellipticity_nearest}
\end{figure}

In contrast to these suggestions, some morphological studies provide results which are consistent with our Ly$\alpha$ dependence. \citet{2007ApJ...656....1L} have investigated morphologies of 216 $z\sim2-3$ LBGs with spectroscopic redshifts using the {\it HST} data. They quantify a multiplicity of a galaxy with the multiplicity parameter $\Psi$. The value of $\Psi$ is zero for a galaxy with a single component, and becomes positive for a galaxy with multiple components. This parameter is used to find multiple components, which is similar to our close-pair method. They reveal that the Ly$\alpha$ EW monotonically increases from $0$ to $15$ \,\AA\, with decreasing $\Psi$. This trend is consistent with the behavior of our merger fractions in the EW range of $20-200$ \,\AA, which is larger than the EW range of \citet{2007ApJ...656....1L}. \citet{2010A&A...514A..64P} have measured the asymmetry parameter for $z\sim3$ LBGs with and without Ly$\alpha$ emission. They find no difference of $A$ between LBGs with a high and a low EW. The decrease in the merger fraction for our LAEs might be found in the wider dynamic range in $EW({\rm Ly\alpha})$ than that of their LBG sample.

Another explanation is that dust created by past star-formation in individual pre-mergers produces the anti-correlation between Ly$\alpha$ EW and merger fraction. The star formation triggered by a major merger would enhance Ly$\alpha$ emission, but Ly$\alpha$ photons could be absorbed by the dust already existing in individual evolved galaxies. For this reason, Ly$\alpha$ EW would be less enhanced in a system consisting of evolved major merger components, which would yield the anti-correlation. 

In addition to observational studies, \citet{2013ApJ...773..151Y} have investigated physical properties of interacting Ly$\alpha$ Blob (LAB) pairs by combining hydrodynamical simulations with three-dimensional radiative transfer calculations. The SFR of LABs is boosted during each galaxy-coalescence phase. In contrast to SFR, Ly$\alpha$ EW fluctuates in the range of $20-100$ \,\AA, and is not enhanced to $>100$\,\AA\, even at the time of coalescences. Note that the simulated LABs have a larger size and slightly-brighter Ly$\alpha$ luminosity than those of our normal LAEs. 

There is a possibility that the anti-correlation between the merger fraction and Ly$\alpha$ EW would be produced by a difference of viewing angle. In this study, we have defined mergers as objects with multi-components or interacting features shown in the plane of the sky. However, mergers along the line-of-sight would boost the radial velocity of the surrounding gas clouds, and consequently enhance the Ly$\alpha$ escape in the direction to the observer. The detection of the line-of-sight mergers could be more challenging than the identification of interacting events shown in the plane of the sky.

\subsection{Where Is Ly$\alpha$ Emitted From?}\label{subsec_discuss_offset}

We find that there are few LAEs with a high Ly$\alpha$ EW and a large $\delta_{\rm Ly\alpha}$. The dependence of $\delta_{\rm Ly\alpha}$ on Ly$\alpha$ EW would suggest that Ly$\alpha$ photons could be heavily attenuated by dust on the long path lengths prior to escaping an H {\sc i} cloud. This might yield the $\delta_{\rm Ly\alpha}$ difference in objects with a high and a low Ly$\alpha$ EW. 

Prior to our statistical study, several NB imaging studies have also estimated $\delta_{\rm Ly\alpha}$ for high-$z$ LAEs. \citet{2013ApJ...773..153J} have studied rest-frame UV morphologies of 51 LAEs at $z\sim5.7$, $6.5$, and $7.0$ using the {\it HST} data. The Ly$\alpha$ positions of these LAEs have been estimated in NB images taken with Subaru/Suprime-Cam. They find that several LAEs show evidence of positional offset between UV and Ly$\alpha$ emission. In these $z\sim6-7$ LAE samples, LAEs with a spatially-symmetric light profile tend to have a small $\delta_{\rm Ly\alpha}$. The offset is also shown in an extended Ly$\alpha$ emission, Himiko, at $z=6.595$ \citep{2013ApJ...778..102O}. \citet{2011MNRAS.418.1115R} find a large positional offset in a galaxy at $z=3.334$ based on a deep spectroscopic survey. For the $z=3.334$ galaxy, the Ly$\alpha$ and UV structure is highly peculiar and is likely to be affected by several physical processes such as the cold gas inflow. 

In contrast, most of these LAEs with a large $\delta_{\rm Ly\alpha}$ show merger and/or interacting features. \citet{2011ApJ...735....5F} have performed high resolution imaging observations with an NB filter on {\it HST} for three LAEs at $z=4.4$. They do not find positional offsets between resolved Ly$\alpha$ and UV continuum emission. All of the three LAEs taken with {\it HST} also show no evidence of major merger/galaxy interactions. 

These results would indicate that the relatively-small $\delta_{\rm Ly\alpha}$ in high EW objects is originated from physically-stable and spatially-symmetric H {\sc i} gas clouds around central ionizing source(s). On the contrary, the large $\delta_{\rm Ly\alpha}$ would result from inhomogeneous H {\sc i} gas clouds disturbed by a merger. The disturbed clouds prevent Ly$\alpha$ radiation from escaping directly along the line of sight, making a large $\delta_{\rm Ly\alpha}$ from an original position of stellar component. A large number of the resonant scattering would suppress Ly$\alpha$ EW on the long path lengths in the disturbed clouds.

\subsection{Galactic Inclination Effect on Ly$\alpha$ Emissivity}\label{subsec_discuss_ellipticity}

We find that there is a trend that the LAEs with a large Ly$\alpha$ EW have a small ellipticity. Figures \ref{fig_ew_ellipticity_brightest} and \ref{fig_ew_ellipticity_nearest} show a possible absence of objects at a high ellipticity and Ly$\alpha$ EW region (in the upper-right corner in the figures). This trend is consistent with the recent theoretical claims that Ly$\alpha$ photons can more easily escape from face-on disks having a small ellipticity, due to a low {\sc Hi} column density \citep[e.g., ][]{2012A&A...546A.111V,2012ApJ...754..118Y}. 

The ellipticity is an useful indicator of the galactic disk inclination. \citet{2012A&A...546A.111V} have investigated quantitatively the effect by using their Ly$\alpha$ radiative transfer code combined with hydrodynamics simulations. They find that Ly$\alpha$ EW strongly depends on the inclination on a simulated galaxy with thick star-forming clouds. From edge-on to face-on, the Ly$\alpha$ EW increases from $-5$ to $90$\,\AA. \citet{2012ApJ...754..118Y} have predicted that the Ly$\alpha$ flux is a hundred times brighter in the face-on direction than the edge-on. 

However, our KS test does not definitively indicate that there is an anti-correlation between Ly$\alpha$ EW and the ellipticity (\S \ref{subsec_result_ellipticity}). The sample size of our high EW LAEs may be too small to obtain conclusive evidence of the anti-correlation. We require a larger LAE sample containing many high EW ($>100$\,\AA) objects bright ($m_{\rm cont}<25$) enough to measure robustly their morphologies. 

\subsection{What is the Physical Origin of Strong Ly$\alpha$ Emission?}\label{subsec_discuss_origin}

Our results of Ly$\alpha$ EW dependence generally support the idea that an H {\sc i} column density is a key quantity determining Ly$\alpha$ emissivity. We find that LAEs with $EW({\rm Ly\alpha})>100$ \,\AA\, tend to be a non-merger (\S \ref{subsec_result_merger}), and compact (Fig. \ref{fig_hlr_ew}), and to have a small ellipticity (\S \ref{subsec_result_ellipticity}) in our structure analyses. Our magnitude cut allows us to fairly compare structural properties between each $EW({\rm Ly\alpha})$ bin under the same ranges of galaxy luminosity correlating mass (Fig. \ref{fig_ew_mag}). We also verify the above trends by using objects with a similar size, but we do not find any significant changes. These results could attribute the $EW({\rm Ly\alpha})$ dependences on the LAE structures to predominantly Ly$\alpha$ emissivity rather than the galaxy mass. These trends do not depend strongly on whether we use the brightest or nearest counterparts.  

Recent spectroscopic studies measure the Ly$\alpha$ velocity offset, $\Delta v_{\rm Ly\alpha}$, from the systemic redshift estimated from nebular lines for a number of LAEs \citep{2013ApJ...765...70H,2014arXiv1402.1168S}. Their kinematic analyses have suggested that LAEs typically have a smaller $\Delta v_{\rm Ly\alpha}$ than that of LBGs with a lower Ly$\alpha$ EW, while their outflowing velocities are similar in the two populations. This indicates that the small $\Delta v_{\rm Ly\alpha}$ of LAEs is caused by a low H {\sc i} column density. On the other hand, NIR spectroscopy by \citet{2013ApJ...769....3N} has suggested that LAEs have a large $[$O {\sc iii}$]/[$O {\sc ii}$]$ ratio, indicating these systems are highly ionized with density-bounded H {\sc ii} regions. This tendency has been confirmed by a subsequent systematic study in \citet{2013arXiv1309.0207N}. The large $[$O {\sc iii}$]/[$O {\sc ii}$]$ ratio also indicates a low column density of H {\sc i} gas. On the basis of these results on the gas distribution and abundances, the difference in H {\sc i} column density simply explains the Ly$\alpha$-EW dependences of  the merger fraction, the Ly$\alpha$ spatial offset, and the galaxy inclination. For objects with density-bounded H {\sc ii} regions, Ly$\alpha$ photons would directly escape from central ionizing sources, which produce a small $\delta_{\rm Ly\alpha}$. The low H {\sc i} abundance along the line of sight also induces the preferential escape of Ly$\alpha$ to the face-on direction. For these reasons, ionized regions with small amounts of H {\sc i} gas would dominate in the subsample of our LAEs with $EW>100$\,\AA.

\section{SUMMARY AND CONCLUSION}\label{sec_conclusion}

We examine the structural properties of LAEs at $z=2.2$ using the {\it HST} high resolution images in order to investigate the Ly$\alpha$ emitting mechanisms. By using the large LAE sample of $426$ objects, we study statistically the Ly$\alpha$-EW dependence on the merger fraction, the Ly$\alpha$ spatial offset, $\delta_{\rm Ly\alpha}$, and ellipticity, for the first time. 

The conclusions of our structure analyses for LAEs are summarized below. 
 
\begin{itemize}
  \item Our results of the merger fraction and the ellipticity distribution are consistent with those in previous morphological studies for LAEs at various redshifts. The merger fraction and the average ellipticity of LAE's stellar component are $10-30$\% and $0.4-0.6$, respectively. 
  \item The merger fractions of LAEs do {\it not} significantly increase with their Ly$\alpha$ EW. This trend is opposite to the physical picture in which the Ly$\alpha$ EW is boosted by the galaxy merger and interaction. H {\sc i} clouds disturbed by merger would envelop a central ionizing source instead of making holes in the gas clouds. The disturbed clouds may not allow Ly$\alpha$ photons to easily escape from a galaxy. 
  \item We successfully identify that some LAEs have a spatial offset between Ly$\alpha$ and stellar-continuum emission peaks by $\sim0.\!\!^{\prime\prime}3-0.\!\!^{\prime\prime}5$ ($\sim 2.5-4$ kpc) beyond our statistical errors. We reveal an anti-correlation between $\delta_{\rm Ly\alpha}$ and $EW({\rm Ly\alpha})$ by KS test with two subsamples of $EW({\rm Ly\alpha})=20-100$ and $>100$\AA. The anti-correlation would suggest that Ly$\alpha$ photons could be heavily attenuated by dust on the long path lengths prior to escaping H {\sc i} clouds. On the contrary, a large $\delta_{\rm Ly\alpha}$ would result from inhomogeneous H {\sc i} gas clouds disturbed by merger. The disturbed clouds prevent Ly$\alpha$ radiation from escaping directly along a line of sight, giving a large $\delta_{\rm Ly\alpha}$. Resonant scattering of long path lengths would suppress Ly$\alpha$ EW in the disturbed clouds. 
  \item We find that there is a trend that LAEs with a large Ly$\alpha$ EW have a small ellipticity. This is consistent with the recent theoretical claims that Ly$\alpha$ photons can more easily escape from face-on disks having a small ellipticity, due to a low {\sc Hi} column density, although our KS test indicates that this trend is not significant in a statistical sense. However, this KS test result might be originated from the small number of bright and spatially-resolved objects with a high EW whose morphological properties are estimated robustly. 
  \item Our results of Ly$\alpha$-EW dependence generally support the idea that an H {\sc i} column density is a key quantity determining Ly$\alpha$ emissivity. In this condition, Ly$\alpha$ photons would directly escape from central ionizing sources. The difference in H {\sc i} abundance along the line of sight is expected to yield naturally the Ly$\alpha$-EW dependences of the merger fraction, the Ly$\alpha$ spatial offset, and the galaxy inclination.
 \end{itemize}

An upcoming extensive survey for LAEs at $z=2-7$ with Hyper Suprime-Cam (HSC) on Subaru will identify a large number of unique high EW objects whose number is not high enough in our study. Future HSC studies will test the possible anti-correlation between Ly$\alpha$ EW and ellipticity with large statistical samples.


\acknowledgments

We would like to thank Anne Verhamme, Zheng Zheng, Michael Rauch, Lucia Guaita, and Akio Inoue for useful discussion, and an anonymous referee for constructive comments. This work is based on observations taken by the CANDELS Multi-Cycle Treasury Program with the NASA/ESA HST, which is operated by the Association of Universities for Research in Astronomy, Inc., under NASA contract NAS5-26555. The NB387 data used in this work are collected at the Subaru Telescope, which is operated by the National Astronomical Observatory of Japan. This work is based in part on observations made with the Spitzer Space Telescope, which is operated by the Jet Propulsion Laboratory, California Institute of Technology under a contract with NASA. Support for this work was provided by NASA through an award issued by JPL/Caltech. This work was supported by World Premier International Research Center Initiative (WPI Initiative), MEXT, Japan. This work was supported by KAKENHI (23244025) and (21244013) Grant-in-Aid for Scientific Research (A) through Japan Society for the Promotion of Science (JSPS). 

{\it Facilities:} \facility{Subaru/Suprime-Cam (NAOJ)}, \facility{HST/ACS, WFC3}.


\bibliographystyle{apj}
\bibliography{reference}

\end{document}